\def\ojet{\Omega_{\rm jet}}

\def\eiso{E_{\rm iso}}
\def\egamma{E_{\gamma}}

\documentclass[
    ,final            
  ]
  {aipproc}

\layoutstyle{6x9}

\begin{document}

\title{A Unified Jet Model of X-Ray Flashes and Gamma-Ray Bursts}

\author{D. Q. Lamb}{
  address={Department of Astronomy \& Astrophysics, University of
Chicago, Chicago, IL 60637}
  ,altaddress={d-lamb@uchicago.edu}
}

\author{T. Q. Donaghy}{
  address={Department of Astronomy \& Astrophysics, University of
Chicago, Chicago, IL 60637}
}

\author{C. Graziani}{
  address={Department of Astronomy \& Astrophysics, University of
Chicago, Chicago, IL 60637}
}

\begin{abstract}

HETE-2 has provided strong evidence that the properties of X-Ray
Flashes (XRFs) and GRBs form a continuum, and therefore that these two
types of bursts are the same phenomenon.  We show that both the
structured jet and the uniform jet models can explain the observed
properties of GRBs reasonably well.  However, if one tries to account
for the properties of both XRFs and GRBs in a unified picture, the
uniform jet model works reasonably well while the structured jet model
fails utterly.  The uniform jet model of XRFs and GRBs implies that
most GRBs have very small jet opening angles ($\sim$ half a degree). 
This suggests that magnetic fields play a crucial role in GRB jets. 
The model also implies that the energy radiated in gamma rays is $\sim$
100 times smaller than has been thought.  Most importantly, the model
implies that there are $\sim 10^4 -10^5$ more bursts with very small
jet opening angles for every such burst we see.  Thus the rate of GRBs
could be comparable to the rate of Type Ic core collapse supernovae. 
Accurate, rapid localizations of many XRFs, leading to identification
of their X-ray and optical afterglows and the determination of their
redshifts, will be required in order to confirm or rule out these
profound implications.  


\end{abstract}

\maketitle

\section{Introduction}

Two-thirds of all HETE-2--localized bursts are either ``X-ray-rich'' or
X-Ray Flashes (XRFs); of these, one-third are XRFs \footnote{We define
``X-ray-rich'' GRBs and XRFs as those events for which $\log
[S_X(2-30~{\rm kev})/S_\gamma(30-400~{\rm kev})] > -0.5$ and 0.0,
respectively.} \citep{sakamoto2003b}.  These events have received
increasing attention in the past several years
\citep{heise2000,kippen2002}, but their nature remains unknown.

XRFs have $t_{90}$ durations between 10 and 200 sec and their sky
distribution is consistent with isotropy.  In these respects, XRFs are
similar to ``classical'' GRBs.  A joint analysis of WFC/BATSE spectral
data showed that the low-energy and high-energy photon indices of XRFs
are $-1$ and $\sim -2.5$, respectively, which are similar to those of
GRBs, but that the XRFs had spectral peak energies $E_{\rm peak}^{\rm
obs}$ that were much lower than those of GRBs \citep{kippen2002}. The
only difference between XRFs and GRBs therefore appears to be that XRFs
have lower $E_{\rm peak}^{\rm obs}$ values.  It has therefore been
suggested that XRFs might represent an extension of the GRB population
to bursts with low peak energies.

Clarifying the nature of XRFs and X-ray-rich GRBs, and their connection
to GRBs, could provide a breakthrough in our understanding of the
prompt emission of GRBs.  Analyzing 42 X-ray-rich GRBs and XRFs seen by
FREGATE and/or the WXM instruments on HETE-2, \cite{sakamoto2003b}
find that the XRFs, the X-ray-rich GRBs, and GRBs form a continuum in
the [$S_\gamma(2-400~{\rm kev}), E^{\rm obs}_{\rm peak}$]-plane (see
Figure 1, left-hand panel).  This result strongly suggests that all of
these events are the same phenomenon.

Furthermore, \cite{lamb2003c} have placed 9 HETE-2 GRBs with known
redshifts and 2 XRFs with known redshifts or strong redshift
constraints in the ($E_{\rm iso}, E_{\rm peak}$)-plane (see Figure 1,
right-hand panel).  Here $E_{\rm iso}$ is the isotropic-equivalent
burst energy and $E_{\rm peak}$ is the energy of the peak of the burst
spectrum, measured in the source frame.  The HETE-2 bursts confirm the
relation between $E_{\rm iso}$ and $E_{\rm peak}$ found by
\cite{amati2002} (see also \cite{lloyd2000}) for GRBs and extend it
down in $E_{\rm iso}$ by a factor of 300.  The fact that XRF 020903,
one of the softest events localized by HETE-2 to date, and XRF 030723,
the most recent XRF localized by HETE-2, lie squarely on this relation
\citep{sakamoto2003a,lamb2003c} provides strong evidence that XRFs and
GRBs are the same phenomenon.  However, additional redshift
determinations are clearly needed for XRFs with 1 keV $< E_{\rm peak} <
30$ keV energy in order to confirm these results.

\begin{figure}[t]
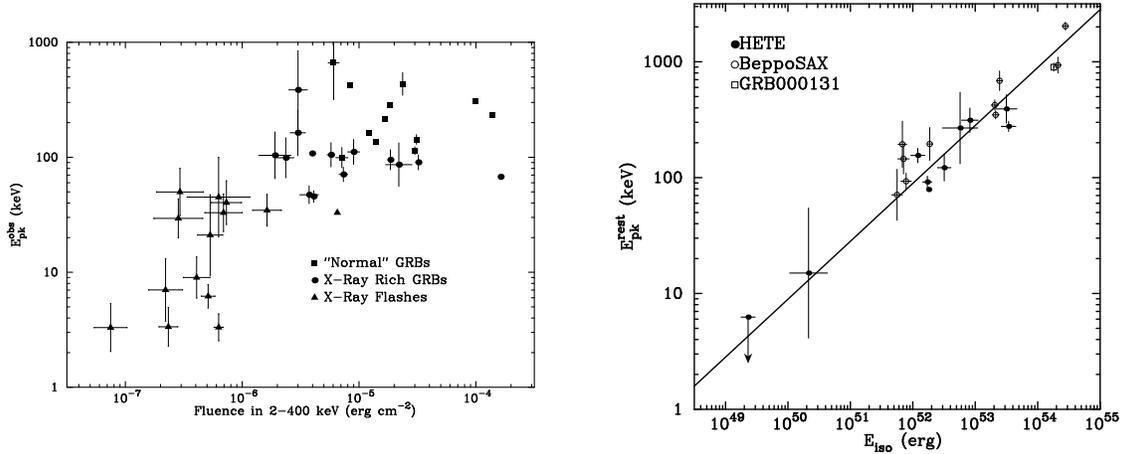

\begin{minipage}{2.1truein}
\includegraphics[width=2.0truein, angle=-90, clip]{Epk-Fluence_by_hardness.bw.ps}
\end{minipage}
\hspace{1.0in}
\begin{minipage}{2.7truein}
\includegraphics[width=2.65truein, clip]{Epk-Erad-HETE+BSAX.bw.ps}
\end{minipage}
\hfill
\caption{Distribution of HETE-2 bursts in the [$S(2-400~{\rm keV}),
E^{\rm obs}_{\rm peak}$]-plane, showing XRFs, X-ray-rich GRBs, and GRBs 
(left panel).    From \cite{sakamoto2003b}.
Distribution of HETE-2 and BeppoSAX bursts in the ($E_{\rm
iso}$,$E_{\rm peak}$)-plane, where $E_{\rm iso}$ and $E_{\rm peak}$ are
the isotropic-equivalent GRB energy and the peak of the GRB spectrum in
the source frame (right panel).   The HETE-2 bursts confirm the relation
between $E_{\rm iso}$ and $E_{\rm peak}$ found by Amati et al. (2002),
and extend it by a factor $\sim 300$ in $E_{\rm iso}$.  The bursts with
the lowest and second-lowest values of $E_{\rm iso}$ are XRFs 020903
and 030723. From \cite{lamb2003c}.}
\label{hete_bsax}
\end{figure}

Figure 2 shows a simulation of the expected distribution of bursts in
the ($E_{\rm iso}, E_{\rm peak}$)-plane (left panel) and in the 
($F^{\rm peak}_N,E_{\rm peak}$)-plane (right panel), assuming that the
\citep{amati2002} relation holds for XRFs as well as for GRBs
\citep{lamb2003}, as is strongly suggested by the HETE-2 results.  The
SXC, WXM, and FREGATE instruments on HETE-2 have thresholds of $1-6$
keV and considerable effective areas in the  X-ray energy range.  Thus
HETE-2 is ideally suited for detecting and studying XRFs.  In contrast,
BAT on {\it Swift} has a nominal threshold of 20 keV.  This simulation
shows that the WXM and SXC instruments on HETE-2 detect many times more
bursts with $ E_{\rm peak} < 10$  keV than will BAT on {\it Swift}.

\section{XRFs as a Probe of GRB Jet Structure, GRB Rate, and Core
Collapse Supernovae}

Most GRBs have a ``standard'' energy
\cite{frail2001,panaitescu2001,bloom2003}; i.e, if their isotropic
equivalent energy is corrected for the jet opening angle inferred from
the jet break time, most GRBs have the same radiated energy, $\egamma =
1.3 \times 10^{51}$ ergs, to within a factor of $\sim$ 2-3.

Two models of GRB jets have received widespread attention:

\begin{itemize}

\item
The ``structured jet'' model (see the left-hand panel of Figure 3). In
this model, all GRBs produce jets with the same structure
\citep{rossi2002,woosley2003,zhang2002,meszaros2002}.  The
isotropic-equivalent energy and luminosity is assumed to decrease as
the viewing angle $\theta_v$ as measured from the jet axis increases. 
The wide range in values of $E_{\rm iso}$ is attributed  to differences
in the viewing angle $\theta_v$.  In order to recover the ``standard
energy'' result \citep{frail2001}, $E_{\rm iso} (\theta_v) \sim
\theta_v^{-2}$ is required \citep{zhang2002}.
\bigskip

\item
The ``uniform jet'' model (see the right-hand panel of Figure 3). In
this model GRBs produce jets with very different jet opening angles 
$\theta_{\rm jet}$.  For $\theta < \theta_{\rm jet}$, $E_{\rm iso}
(\theta_v)$ = constant while for $\theta > \theta_{\rm jet}$, $E_{\rm
iso} (\theta_v) = 0$.

\end{itemize}

\begin{figure}[t]
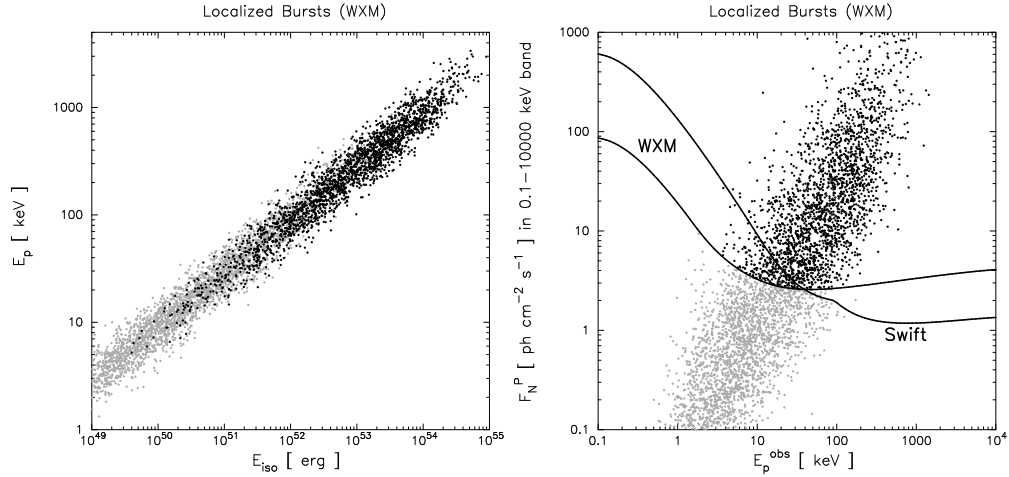

\includegraphics[height=.3\textheight,angle=270]{Eiso_Ep.m2.bw.ps}
\includegraphics[height=.3\textheight,angle=270]{Ep_obs_FNP.hete.m2.bw.ps}
\caption{Expected distribution of bursts in the ($E_{\rm iso}, E_{\rm
peak}$)-plane (left panel) and in the  ($F^{\rm peak}_N,E_{\rm
peak}$)-plane (right panel), assuming that the Amati et al. (2002)
relation holds for XRFs as well as for GRBs, as strongly suggested by
the HETE-2 results.  Black dots are simulated bursts that the WXM on
HETE-2 detects; gray dots are simulated bursts that it does not detect. 
The curved lines in the right-hand panel show the
threshold sensitivities of the WXM on HETE-2 and BAT on Swift.  From
\cite{lamb2003}.}
\label{model_dotplots}
\end{figure}

As we have seen, HETE-2 has provided strong evidence that the
properties of XRFs, X-ray-rich GRBs, and GRBs form a continuum, and
that these bursts are therefore the same phenomenon.  If this is true,
it immediately implies that the $\egamma$ inferred by 
\citep{frail2001} is too large by a factor of at least 100
\citep{lamb2003}.  The reason is that the values of $E_{\rm iso}$ for
XRF 020903 \citep{sakamoto2003a} and XRF 030723 \citep{lamb2003c} are
$\sim$ 100 times smaller than the value of  $\egamma$ inferred by Frail
et al. -- an impossibility.

HETE-2 has also provided strong evidence that, in going from XRFs to 
GRBs, $E_{\rm iso}$ changes by a factor $\sim 10^5$ (see Figure 1,
right-hand panel).  If one tries to explain only the range in $E_{\rm
iso}$ corresponding to GRBs, both the uniform jet model and the
structured jet model work reasonably well.  However, if one tries to
explain the range in $E_{\rm iso}$ of a factor $\sim 10^5$ that is
required in order to accommodate both XRFs and GRBs in a unified
description, the uniform jet works reasonably well while the structured
jet model does not.

\begin{figure}[t]
\includegraphics[height=.3\textheight]{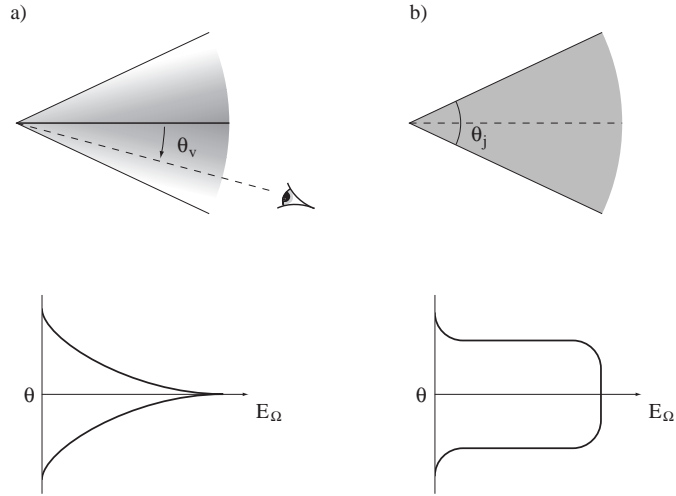}
\caption{Schematic diagrams of universal jet model and jet model of GRBs
\citep{ramirez-ruiz2002}. 
In the universal jet model, the isotropic-equivalent energy and
luminosity is assumed to decrease as the viewing angle $\theta_v$ as
measured from the jet axis increases.  In order to recover the
``standard energy'' result \citep{frail2001}, $E_{\rm iso} (\theta_v) \sim
\theta_v^{-2}$ is required.  In the uniform jet model, GRBs produce
jets with a large range of jet opening angles  $\theta_{\rm jet}$.  For
$\theta < \theta_{\rm jet}$, $E_{\rm iso} (\theta_v)$ = constant while
for $\theta > \theta_{\rm jet}$, $E_{\rm iso} (\theta_v) = 0$.}
\label{jet_schematic}
\end{figure}

\begin{figure}[h]
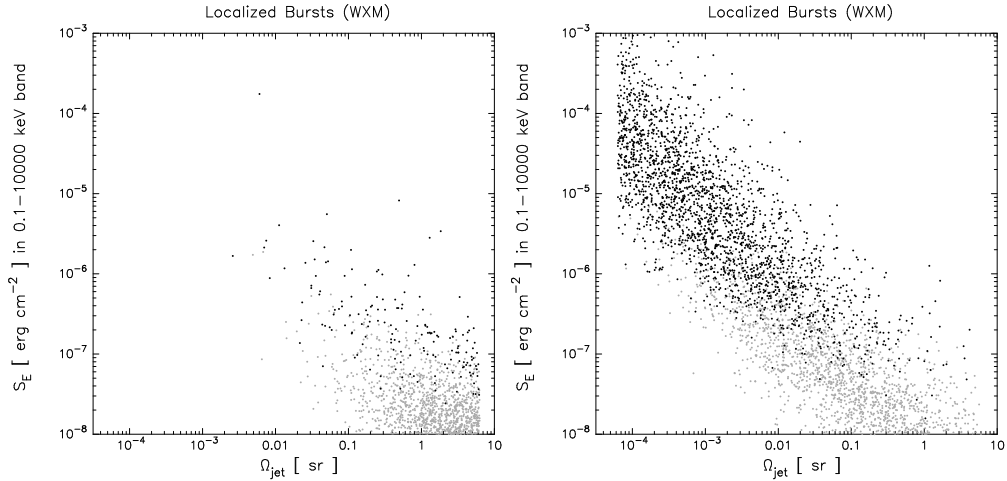

\includegraphics[height=.3\textheight,angle=270]{omega_SE.univ1m.bw.ps}
\includegraphics[height=.3\textheight,angle=270]{omega_SE.m2.bw.ps}
\caption{Expected distribution of bursts in the ($\Omega_{\rm
jet},S_E$)-plane for the universal jet model (left panel) and uniform
jet model (right panel), assuming that the Amati et al. (2002) relation
holds for XRFs as well as for GRBs, as the HETE-2 results strongly
suggest.  From \cite{lamb2003}.}
\label{omega_plots}
\end{figure}

\begin{figure}[htb]
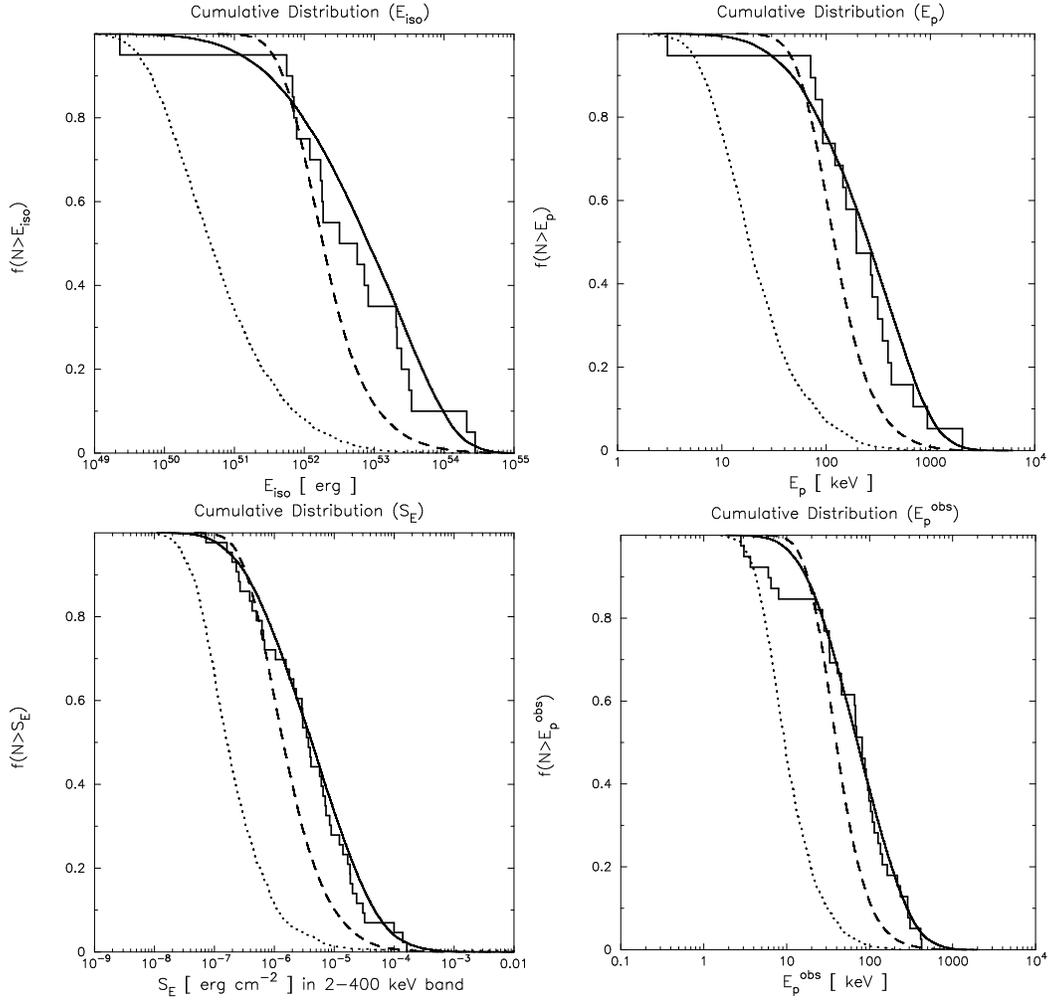

\begin{minipage}{2.7in}
\includegraphics[width=2.6in,angle=270]{both.Eiso.comp.cuml.bw.ps}
\\
\includegraphics[width=2.6in,angle=270]{hete.flu.2_400.comp.cuml.bw.ps}
\end{minipage}
\begin{minipage}{2.7in}
\includegraphics[width=2.6in,angle=270]{both.Ep.comp.cuml.bw.ps}
\\
\includegraphics[width=2.6in,angle=270]{hete.Ep_obs.comp.cuml.bw.ps}
\end{minipage}
\caption{Top row: cumulative distributions of $E_{\rm iso}$ (left
panel) and $E_{\rm peak}$ (right panel) predicted by various models,
compared to the observed cumulative distributions of these quantities. 
Bottom row: cumulative distributions of $S(2-400 {\rm keV})$ (left
panel) and $E^{\rm obs}_{\rm peak}$ (right panel) predicted by various
models, compared to the observed cumulative distributions of these
quantities.  The uniform jet model is shown as a solid line.  The
cumulative distributions corresponding to the best-fit structured jet
model that explains XRFs and GRBs are shown as dotted lines; the
cumulative distributions corresponding to the best-fit structured jet
model that explains GRBs alone are shown as dashed lines.  The
structured jet model provides a reasonable fit to GRBs alone but cannot
provide a unified picture of both XRFs and GRBs, whereas the uniform
jet model can.  From \cite{lamb2003}.}
\label{cuml_dist}
\end{figure}

The reason is the following:  the observational implications of the
structured jet model and the uniform jet model differ dramatically if
they are required to explain XRFs and GRBs in a unified picture.  In
the structured jet model, most viewing angles $\theta_v$ are $\approx
90^\circ$.  This implies that the number of XRFs should exceed the
number of GRBs by many orders of magnitude, something that HETE-2 does
not observe (see Figures 1, 2, 4, and 5).  On the other hand, by
choosing $N(\Omega_{\rm jet}) \sim \Omega_{\rm jet}^{-2}$, the uniform
jet model predicts equal numbers of bursts per logarithmic decade in
$E_{\rm iso}$ (and $S_E$), which is exactly what HETE-2 sees (again,
see Figures 1, 2, 4, and 5) \citep{lamb2003}.
Thus, if $E_{\rm iso}$ spans a range $\sim 10^5$, as the HETE-2 results
strongly suggest, the uniform jet model can provide a unified picture
of both XRFs and GRBs, whereas the structured jet model cannot.  This
means that XRFs provide a powerful probe of GRB jet structure.

A range in $\eiso$ of $10^5$, which is what the HETE-2 results strongly
suggest, requires a {\it minimum} range in $\Delta \ojet$ of $10^4 -
10^5$ in the uniform jet model.  Thus the unified picture of XRFs and
GRBs in the uniform jet model implies that there are $\sim 10^4 - 10^5$
more bursts with very small $\ojet$'s for every such burst we see;
i.e., the rate of GRBs may be $\sim 100$ times greater than has
been thought.

In addition, since the observed ratio of the rate of Type Ic SNe to the
rate of GRBs in the observable universe is $R_{\rm Type\ Ic}/ R_{\rm
GRB} \sim 10^5$ \citep{lamb1999}, a unified picture of XRFs and GRBs in
the uniform jet model implies that the GRB rate is comparable to
that of Type Ic SNe \citep{lamb2003}.  More spherically
symmetric jets yield XRFs and narrow jets produce GRBs.  Thus XRFs and
GRBs provide a combination of GRB/SN samples that would enable
astronomers to study the relationship between the degree of jet-like
behavior of the GRB and the properties of the supernova (brightness,
polarization $\Leftrightarrow$ asphericity of the explosion, velocity
of the explosion $\Leftrightarrow$ kinetic energy of the explosion,
etc.).  GRBs may therefore provide a unique laboratory for
understanding Type Ic core collapse supernovae.

A unified picture of XRFs and GRBs in the uniform jet model also implies
that many Type Ic SNe produce narrow jets, which may suggest that the
collapsing cores of many Type Ic supernovae are rapidly rotating. 
Finally, such a unified picture implies that the total radiated energy in
gamma rays $E_\gamma$ is $\sim$ 100 times smaller than has been thought 
\cite{lamb2003}.

\end{document}